\documentclass[]{zakoproc}
\usepackage{graphicx}
\usepackage{wrapfig}
\begin{document}

\title{Interaction of Fanaroff-Riley class II radio jets with a randomly magnetised intra-cluster medium}
\author{Mart\'{\i}n Huarte-Espinosa\footnote{martinhe@pas.rochester.edu}$^{~,1,2,5}$, Martin Krause$^{3,4}$ \\
\& Paul Alexander$^{2,5}$}
\institute{
$^1$Department of Physics and Astronomy, University of Rochester,
NY.  $^2$Astrophysics Group, Cavendish Laboratory, University of
Cambridge, UK.  $^3$Universit\"atssternwarte M\"unchen, Germany.
$^4$Max-Planck-Institut f\"ur Extraterrestrische Physik, Garching,
Germany.  $^5$Kavli Institute for Cosmology Cambridge, Cambridge,
UK.
}

\markboth{M. Huarte-Espinosa}{Interaction of FR~II radio jets with a magnetized ICM}

\maketitle

\begin{abstract}
We present 3-D MHD and synthetic numerical simulations to follow
the evolution of randomly magnetized intra-cluster medium plasma
under the effects of powerful, light, hypersonic and bipolar jets.
We prescribe the cluster magnetic field (CMF) as a Gaussian random
field with power law energy spectrum tuned to the expectation for
Kolmogorov turbulence.  We investigate the power of jets and the
viewing angle used for the synthetic Rotation Measure (RM) observations.
We find the model radio sources introduce and amplify fluctuations
on the RM statistical properties; the average RM and the RM standard
deviation are increased by the action of the jets.  This may lead
to overestimations of the CMFs’ strength up to 70\%.  The effect
correlates with the jet power. Jets distort and amplify
CMFs especially near the edges of the lobes and the jets’ heads.
Thus the RM structure functions are flattened at scales comparable
to the source size. Jet-produced RM enhancements depend on the orientation 
of the jet axis to the line of sight and are thus more apparent in 
quasars than in radio galaxies.
\end{abstract}

\section{Introduction}
\vskip-.35cm
The Faraday rotation effect is observed 
against the polarised radio emission of extragalactic radio sources
traveling through the ICM, revealing magnetic fields
of $\sim\,$100\,kpc~scale threading this media.  RM maps
indicate that cluster magnetic fields (CMFs) 
have a random structure with more power on larger scales, 
their strength seems to be of order $\mu$G and
proportional to both the ICM density distribution 
and the cluster cooling
flow related accretion rates (Carilli \& Taylor~\cite{r1}).
The origin, evolution and role of CMFs on the ICM stability are
open questions. Also, since AGN jets have strong effects on the
ICM, it is not clear to what extent, and how, they affect both the
CMFs and their RM characterisation.  We investigate this in
Huarte-Espinosa, Krause \& Alexander~\cite{r2}.

\section{Model}
\vskip-.35cm
Using Flash~3.1 (Fryxell et al.~\cite{r2}) we solve the equations of MHD 
with a constrained transport scheme (Lee \& Deaane~\cite{r4})
in a cubic Cartesian domain with \hbox{200$^3$\,cells}. 
The ICM is implemented as a monoatomic ideal gas ($\gamma=5/3$) with
a density profile of $\rho_{\mathrm{ICM}} = \rho_0 (1+(r/a_0)^2)^{-1}$
in magnetohydrostatic equilibrium with central gravity. Magnetic fields 
in the grid are set up with a Kolmogorov-like structure (following Murgia 
et al.~\cite{r5}), and the thermal over magnetic pressure ratio is of order
unity everywhere. The plasma relaxes for one crossing time and then we inject mass and 
$x$-momentum into a central control cylinder, in a back-to-back fashion. 
We experiment with the jets' 
power using velocities of 40, 80~and 130~Mach, and densities of
0.02$\rho_0$ and 0.004$\rho_0$.

\section{Results}
\vskip-.35cm
We calculate  RM$=$ 812 $\int_0^{\mathrm{D/kpc}} 
   ( n_e / \mathrm{cm}^{-3}         ) 
   ( B_{\parallel} / \mu \,\mathrm{G} ) \, dl 
   \,$ rad\,m$^{-2}$ 
from the jets' cavity contact discontinuity to the end of the domain,
along different viewing angles. We do this at different times, with
and without the jets to assess their effects on the CMFs. The fields
in the region between the cocoon and the bow shock are compressed,
stretched and amplified. In Figure~1 we show the case of jets'
velocity and density of 130~Mach and 0.004$\rho_0$, respectively, 
at a viewing angle of 45$^{\circ}$.
We see that the jets with Mach$=\,$\{80, 130\} are able to 
to increase the ICM magnetic energy energy in proportion to the jet velocity.  
Though the RM structure functions show and preserve the CMFs initial
condition, they are flattened by the jets at scales
of order tens of kpc. 
This scale is larger for sources with fat
cocoons, 
\begin{wrapfigure}[12]{r}{0.63\textwidth}
\vspace{-13pt}
\begin{center}
   \includegraphics[width=.292\textwidth,bb =0.2in 0.2in 6.55in 5.93in,clip=]
{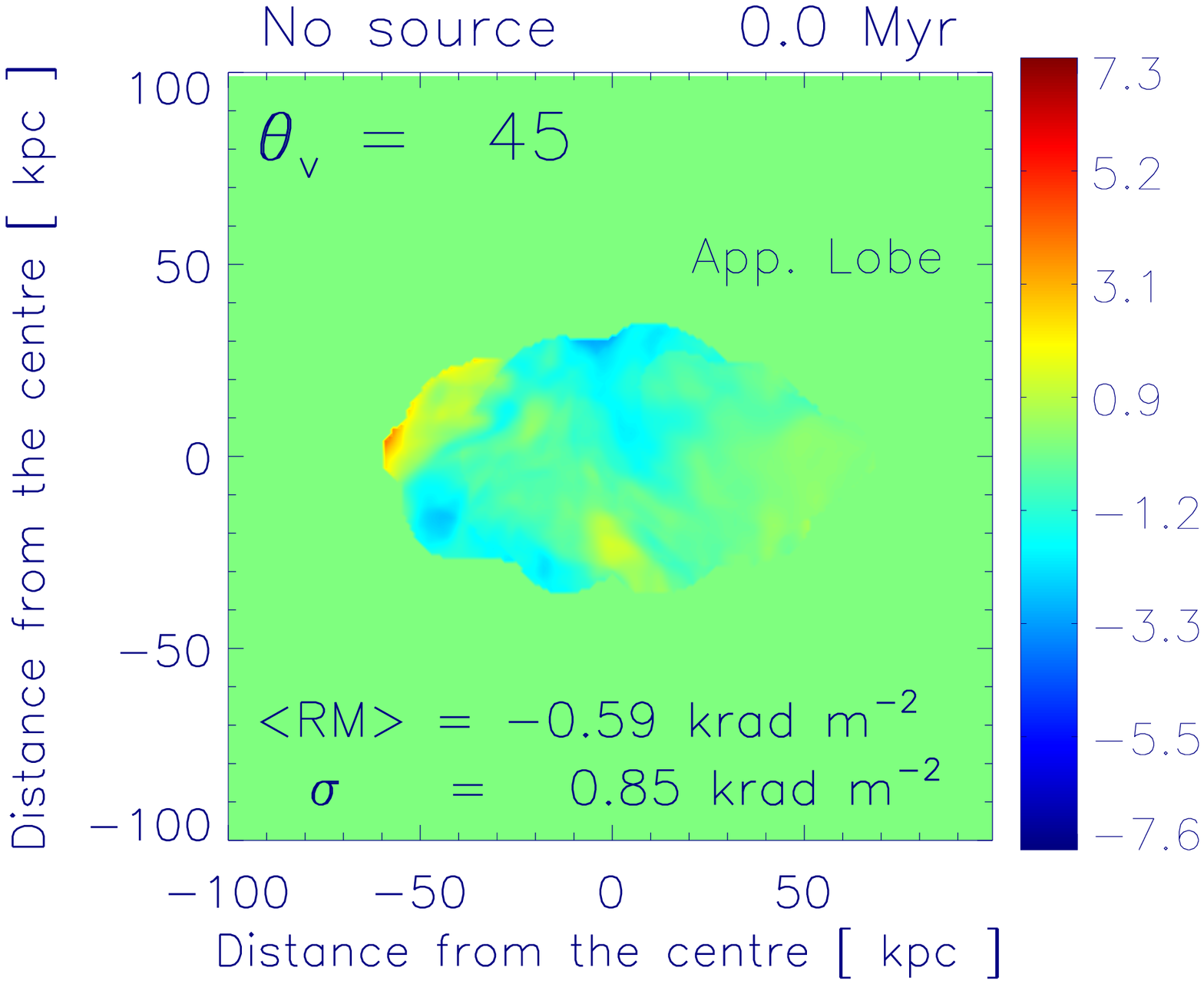}
   \includegraphics[width=.26\textwidth,bb =.9in 0.2in 6.55in 5.93in,clip=]
{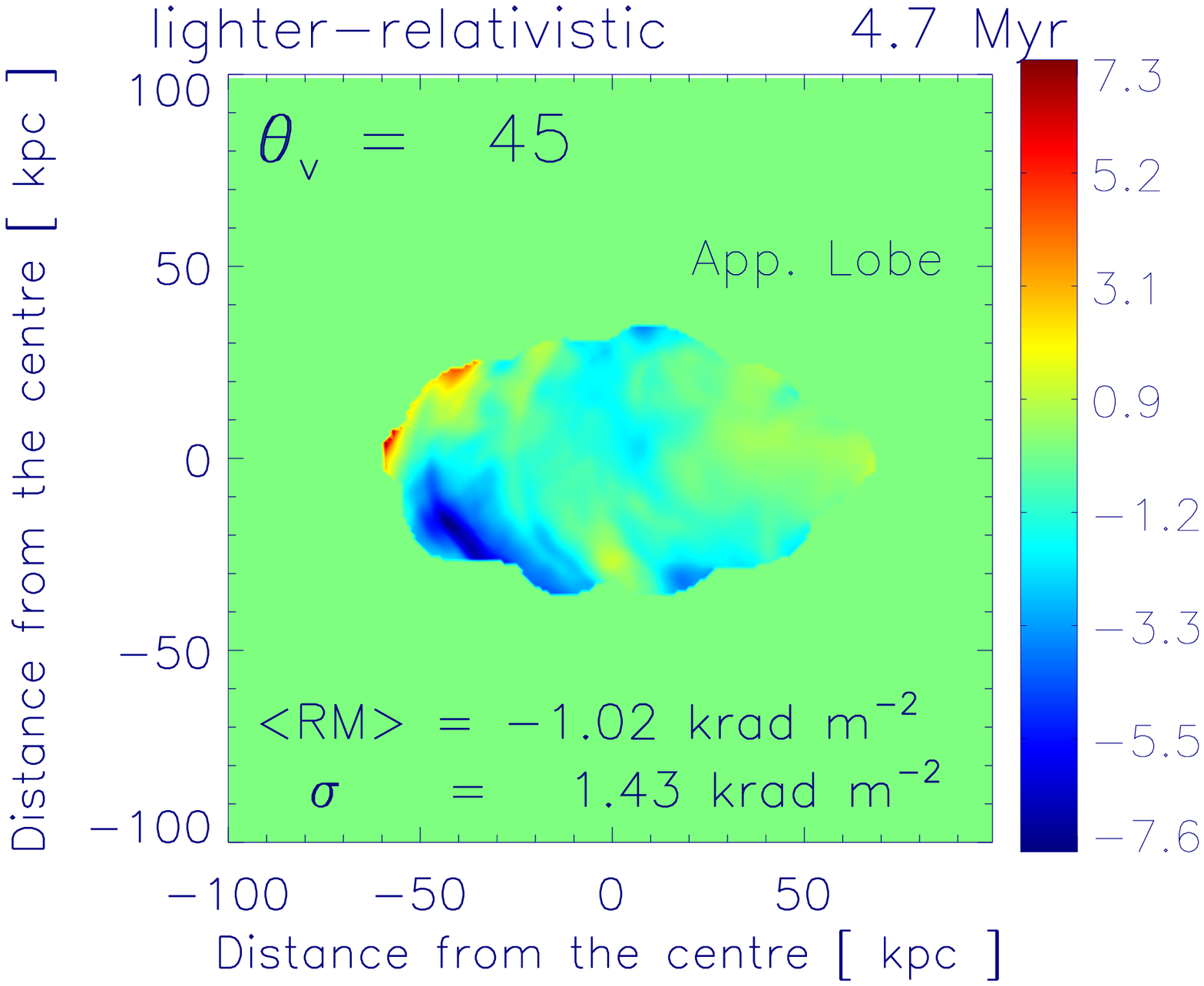}
   \includegraphics[width=.056\textwidth,bb =6.55in 0.2in 7.8in 6.05in,clip=]
{RMmap-05-03-45-15dec09.ps}
\end{center}
\vspace{-20pt}
\caption{{\small Synthetic RM maps for jets with
velocity and density of 130~Mach and 0.004$\rho_0$, respectively,
at a viewing angle of 45$^{\circ}$. Left: before the dynamical effect of jets.
Right: after the dynamical effect of jets.}}
\vspace{-0pt}
\end{wrapfigure}
i.e. light densities.  
We calculate the mean RM for the
approaching and the receding radio lobes vs. the viewing angle.
Intrinsically, the depolarisation is always higher for the receding
lobe; i.e. the Laing-Garrington effect (Laing~\cite{r6}).  This
however is only moderately affected by the radio source expansion,
in such a way that the associated trends tend to be amplified.

\section{Conclusions}
\vskip-.35cm
The jets distort and amplify the CMFs, especially near the edges
of the lobes and the jets' heads.  $\left< \right. $RM$\left.
\right>$ and $\sigma_{RM}$ increase in proportion to the jets'
power.  The effect may lead to overestimations of the CMFs' strength
by about 70\%.  A flattening of the RM structure functions is
produced by the jets, at scales comparable to the source size.
Jet-produced RM enhancements are more apparent in quasars than in
radio galaxies.
\acknowledgements{\small The software used in these investigations was in
part developed by the DOE-supported ASC / Alliance Center for
Astrophysical Thermonuclear Flashes at the University of Chicago.
MHE acknowledges financial support from The Mexican National Council
of Science and Technology, 196898/217314; Dongwook~Lee for the
3D-USM-MHD solver of Flash~3.1.}

\end{document}